\documentclass{article}
\usepackage{spconf,amsmath,graphicx}
\usepackage{comment}
\usepackage{booktabs}
\usepackage{amssymb}
\usepackage{multirow}
\usepackage{newtxtext,newtxmath}

\usepackage{hyperref}
\usepackage{cite}
\hypersetup{
    colorlinks = false,
}

\renewcommand{\Vec}[1]{\textrm{\boldmath $#1$}} 

\title{Virtuoso: Massive Multilingual Speech-Text Joint \\
Semi-Supervised Learning for Text-To-Speech}
\name{
\emph{
Takaaki~Saeki$^{\, 1 \, 3 \, \text{*}}$,\thanks{$^{*}$This work was carried out as an intern at Google, Japan in 2022.}~~
Heiga~Zen$^{\, 1}$,~~
Zhehuai~Chen$^{\, 2}$,~~
Nobuyuki~Morioka$^{\, 1}$,~~
Gary~Wang$^{\, 2}$,
}
\\
\emph{
Yu~Zhang$^{\, 2}$,~~
Ankur~Bapna$^{\, 2}$,~~
Andrew~Rosenberg$^{\, 2}$,~~
Bhuvana~Ramabhadran$^{\, 2}$
}
}
\address{$^1\,$Google, Japan \quad $^2\,$Google, USA \quad  $^3\,$The University of Tokyo, Japan
\\[1mm]
\texttt{\small 
takaaki\_saeki@ipc.i.u-tokyo.ac.jp,  
\{heigazen,zhehuai\}@google.com
}
}
\begin{document}
\ninept
\setlength{\abovedisplayskip}{6pt}
\setlength{\belowdisplayskip}{6pt}
\maketitle
\begin{abstract}
\vspace{-1mm}
This paper proposes \emph{Virtuoso}, a massively multilingual speech--text joint semi-supervised learning framework for text-to-speech synthesis (TTS) models.
Existing multilingual TTS typically supports tens of languages, which are a small fraction of the thousands of languages in the world.
One difficulty to scale multilingual TTS to hundreds of languages is collecting high-quality speech--text paired data in low-resource languages.
This study extends \emph{Maestro}, a speech--text joint pretraining framework for automatic speech recognition (ASR), to speech generation tasks.
To train a TTS model from various types of speech and text data, different training schemes are designed to handle supervised (paired TTS and ASR data) and unsupervised (untranscribed speech and unspoken text) datasets.
Experimental evaluation shows that 1) multilingual TTS models trained on Virtuoso can achieve significantly better naturalness and intelligibility than baseline ones in seen languages, and 2) they can synthesize reasonably intelligible and naturally sounding speech for unseen languages where no high-quality paired TTS data is available.

\end{abstract}
\begin{keywords}
Multilingual text-to-speech synthesis, massive multilingual pretraining, speech--text semi-supervised joint learning.
\end{keywords}
\section{Introduction}\label{sec:intro}

\begin{figure*}
    \centering
    \includegraphics[width=0.98\linewidth, clip]{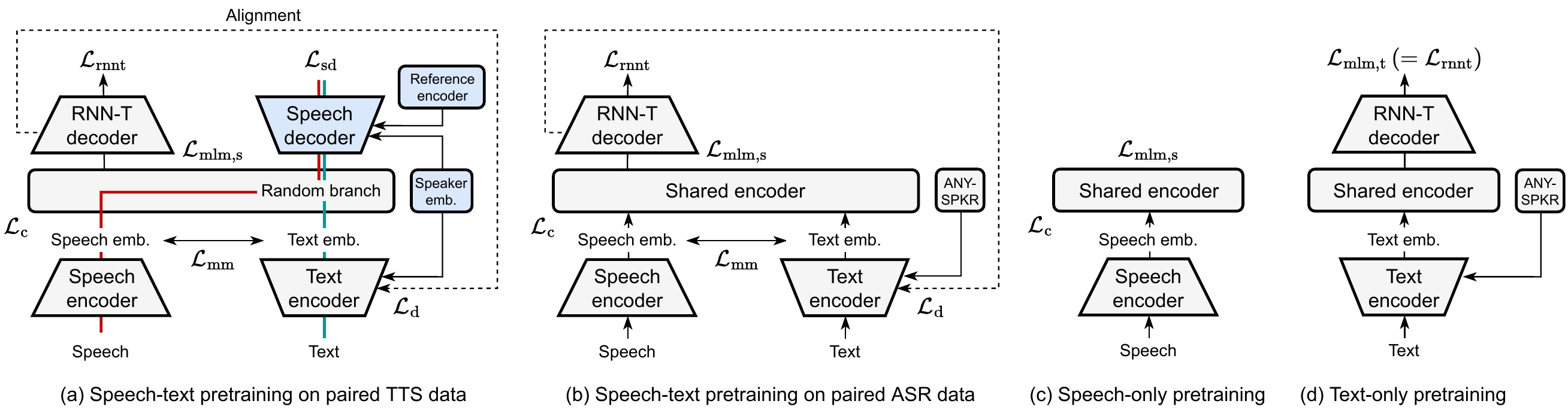}
        \vspace{-1mm}
    \caption{Illustration of Virtuoso framework. (a)~Whole architecture of Virtuoso and its supervised pretraining with paired TTS data where speech features are decoded with speaker labels and reference encoder output. (b)~Supervised pretraining with paired ASR data. (c)~Self-supervised pretraining with untranscribed speech. (d)~Self-supervised pretraining with unspoken text. (b)--(d) follow those in Maestro.}
    \label{fig:method}
    \vspace{-1mm}
\end{figure*}

  \vspace{-1mm}
With the remarkable progress of neural text-to-speech synthesis (TTS) methods, current multilingual TTS systems can synthesize human-like high-quality speech in multiple languages.
Early work on multilingual TTS focused on building a TTS system for rich-resource languages.
For example, Zen et al. \cite{Zen_SLF_TASLP} built a multilingual HMM-based statistical parametric speech synthesis (SPSS) from five Western European languages, and Li and Zen \cite{Li2016MultiLanguageMA} developed a neural network-based multilingual SPSS from six Western European languages.
Recently, the research community has started scaling multilingual TTS to tens of languages.
He et al. \cite{He2021MultilingualBM} proposed a multilingual Byte2Speech TTS model, where 900-hour speech data of 43 languages was used.
However, scaling it to hundreds of languages is still highly challenging due to the difficulty in collecting a large amount of high-quality paired TTS data for low-resource languages  \cite{He2021MultilingualBM}.
To cover thousands of languages, this paper aims to develop a technology that can scale multilingual TTS to hundreds of languages by using diverse speech and text data.

Semi-supervised and self-supervised learning has shown effectiveness for a wide range of speech and natural language processing tasks.
Massive multilingual speech pretraining \cite{Conneau22xlsr} has shown remarkable performance for downstream speech recognition tasks such as multilingual ASR and speech translation. 
Recently, it has been extended to multimodal speech--text joint pretraining \cite{Bapna22mslam,Chen2022MAESTROMS} using speech-text pairs, untranscribed speech, and unspoken text.
Although various approaches of massively multilingual self/semi-supervised learning have been attempted for speech recognition tasks, they have not been fully explored for multilingual speech generation tasks.

This paper proposes \emph{Virtuoso}, a massive multilingual speech--text joint pretraining framework based on self-supervised and semi-supervised learning.
It extends \emph{Maestro} \cite{Chen2022MAESTROMS}, a speech--text semi-supervised pretraining framework for ASR, to speech generation tasks.
Virtuoso allows us to pretrain a multilingual TTS model using unsupervised (untranscribed speech and unspoken text) and supervised (paired TTS and ASR data) datasets with training schemes designed for them, which will allow the model to scale to hundreds of languages.
This work has the following contributions:
\begin{itemize} \leftskip -5.5mm \itemsep -0.5mm
    \item Proposing massive multilingual semi-supervised pretraining for TTS. It leverages different training schemes for ``paired ASR'', ``paired TTS'', ``untranscribed speech'' and ``unspoken text'' data, to train a single TTS model.
    \item Zero-shot TTS, where decent-quality TTS can be achieved for languages not included in the ``paired TTS'' data. 
\end{itemize}

\section{Related Work}\label{sec:related}
Large-scale self-/semi-supervised speech pretraining has been actively studied and applied to various downstream recognition tasks. 
In addition to speech-only pretraining \cite{Baevski2020wav2vec2A,Chung2021w2vBERTCC,Chen2022WavLMLS}, there are multimodal approaches such as TTS-based text injection \cite{chen21injecting} and speech--text joint pretraining \cite{Bapna22mslam,tang-etal-2022-unified,pmlr-v162-bai22d,ao-etal-2022-speecht5}.
Maestro \cite{Chen2022MAESTROMS} performs the modality matching of speech and text embedding to learn speech-aware text representation and vice versa. 
Virtuoso extends Maestro to speech generation tasks by adding a speech decoder on Maestro's shared encoder.

There have been prior studies on joint training of ASR and TTS to improve ASR \cite{Karita_joint_ASR_TTS,Wang2020ImprovingSR}, to obtain alignments \cite{Lim2020JDITJT}, and to scale ASR for low-resource settings \cite{Ren2019AlmostUT,makishima22_interspeech}.
Virtuoso also jointly learns ASR and TTS models, where its shared encoder learns speech--text representation for both recognition and generation tasks.

While most of the existing studies on multilingual TTS \cite{oshaughnessy-1998-multilingual,Li2016MultiLanguageMA,Zhang2019LearningTS,Zhao2020bilingual,Yang2022Lifelong} have focused on a limited number of rich-resource languages, some studies have investigated low-resource languages \cite{prakash19_ssw,ogayo22_interspeech}.
Some previous work has used a byte sequence \cite{Li2019BytesAA,He2021MultilingualBM} as input text tokens to eliminate per-language modules for phoneme inputs and to learn linguistic representations shared across multiple languages.
The prior work which is most similar to this paper is Byte2Speech \cite{He2021MultilingualBM}, where a multilingual TTS model mapping a byte sequence to mel-spectrogram was trained from 900 hours of paired TTS data including 43 languages by 109 speakers.
Virtuoso also uses graphemes or bytes as text input, whereas Virtuoso can use both paired and unpaired data by leveraging semi-supervised learning.

  \vspace{-1mm}
\section{Virtuoso}\label{sec:method}

\vspace{-1mm}
\subsection{Architecture}\label{sec:method-arch}
Fig.~\ref{fig:method}(a) shows the architecture of Virtuoso, consisting of a speech encoder, text encoder, shared encoder, RNN-T decoder, and speech decoder.
It extends Maestro~\cite{Chen2022MAESTROMS} by introducing the speech decoder to learn representation for speech generation tasks.
%
%
The speech decoder predicts speech features with the decoder architecture used in Parallel Tacotron~\cite{elias2021parallel}, given outputs from the shared encoder, a global reference encoder \cite{hsu2018hierarchical}), and speaker embedding.

Like Maestro, Virtuoso can use different types of text input, such as phonemes, graphemes, and bytes \cite{Chen2022MAESTROMS}\footnote{In Maestro, the grapheme input showed the comparable performance to the phoneme, and the byte input improved the zero-resource performance.}.
Although we can derive models for some downstream tasks from Virtuoso even without fine-tuning, e.g., TTS: text encoder$\rightarrow$shared encoder$\rightarrow$speech decoder, ASR: speech encoder$\rightarrow$shared encoder$\rightarrow$RNN-T decoder, and voice conversion: speech encoder$\rightarrow$shared encoder$\rightarrow$speech decoder, we can still fine-tune the model for these tasks.

\vspace{-1mm}
\subsection{Training objectives for supervised and unsupervised data}

We aim to use both supervised (paired TTS and ASR) and unsupervised (untranscribed speech and unspoken text) data to train Virtuoso.
The paired ASR data often contains background noise and channel distortions, whereas the paired TTS data is usually high-quality and recorded in a well-designed environment \cite{Yamagishi2019CSTRVC}.
Also, speaker labels are often available for the paired TTS data, but not for other data types.
To handle such differences in the data nature, Virtuoso uses different training objectives for each of them.

The training losses in Virtuoso include 
speech reconstruction loss $\mathcal{L}_{\text{sd}}$ (to reconstruct speech features from shared representation),
ASR loss $\mathcal{L}_{\text{rnnt}}$ \cite{Bapna22mslam,Chen2022MAESTROMS} (to decode text tokens from shared representation),
contrastive loss $\mathcal{L}_{\text{c}}$ and masked language modeling (MLM) loss $\mathcal{L}_{\text{mlm,s}}$ for speech embedding \cite{Chung2021w2vBERTCC} (to learn self-supervised representation for speech),
aligned MLM loss $\mathcal{L}_{\text{mlm,t}}$ for text embedding \cite{Chen2022MAESTROMS}, 
duration loss $\mathcal{L}_{\text{d}}$ \cite{Chen2022MAESTROMS} (to learn a token duration predictor),
and modality matching (MM) loss $\mathcal{L}_{\text{mm}}$ \cite{Chen2022MAESTROMS} (to unify speech and text embeddings).
For $\mathcal{L}_{\text{sd}}$, we use the iterative loss~\cite{elias2021parallel}, which computes the $L_1$ spectrogram loss at each of the lightweight convolutional blocks in the speech decoder.
%

  \vspace{-1mm}
\subsubsection{Paired TTS data}

Figure~\ref{fig:method}(a) also illustrates the training objective for the paired TTS data.
The loss for the paired TTS data $\mathcal{L}_{\text{tts}}$ is designed as
\begin{equation}
\begin{split}
    \mathcal{L}_{\text{tts}} &= 
        \lambda_{\text{sd}}\mathcal{L}_{\text{sd}} + 
        \lambda_{\text{rnnt}}\mathcal{L}_{\text{rnnt}} + 
        \lambda_{\text{c}}\mathcal{L}_{\text{c}} \nonumber\\
        &+ \lambda_{\text{mlm},\text{s}}\mathcal{L}_{\text{mlm},\text{s}} +
        \lambda_{\text{d}}\mathcal{L}_{\text{d}} +
        \lambda_{\text{mm}}\mathcal{L}_{\text{mm}}\nonumber,
\end{split}
\end{equation}
where $\lambda$ denotes the weighting term of each objective.
The speech decoder takes speaker embedding, a global reference encoder output, and shared encoder outputs then predicts speech features.
Let the shared encoder output and speaker embedding be $\Vec{e}_{\text{shared}}$ and $\Vec{e}_{\text{spkr}}$, respectively.
The predicted speech features $\hat{\Vec{s}}$ is given as $\hat{\Vec{s}} = \theta_{sd}(\Vec{e}_{\text{shared}}, \Vec{e}_{\text{spkr}}, \theta_{\text{ref}}(\Vec{s}))$, where $\theta_{\text{ref}}$ denotes the reference encoder and $\Vec{s}$ is the target speech features.

The shared encoder inputs can be either speech embeddings from the speech encoder or up-sampled text embeddings from the text encoder. 
We denote the former as ``speech branch'' and the latter as ``text branch'', where the former can be viewed as a masked autoencoder~\cite{He2022MaskedAA} and the latter is TTS.
Although we can train the TTS model only with the text branch, a preliminary experiment showed that randomly switching between these branches during training helped the model to converge.

  \vspace{-1mm}
\subsubsection{Paired ASR data}\label{sec:method-obj-asr}

Figure~\ref{fig:method}(b) describes the training objective for the paired ASR data, which is identical to that in Maestro \cite{Chen2022MAESTROMS}.
The objective for the ASR data $\mathcal{L}_{\text{asr}}$ is given as
\begin{equation}
    \mathcal{L}_{\text{asr}} =
        \lambda_{\text{rnnt}}\mathcal{L}_{\text{rnnt}} + 
        \lambda_{\text{c}}\mathcal{L}_{\text{c}} + 
        \lambda_{\text{mlm},\text{s}}\mathcal{L}_{\text{mlm},\text{s}} + 
        \lambda_{\text{d}}\mathcal{L}_{\text{d}} +
        \lambda_{\text{mm}}\mathcal{L}_{\text{mm}}.\nonumber
\end{equation}
Note that we applied masking for speech and text embedding in the same manner as Maestro.
As speaker labels are not available in the paired ASR data, $\mathcal{L}_{\text{d}}$
is computed with a speaker embedding for the wildcard identifier \textsf{ANY-SPKR}.

  \vspace{-1mm}
\subsubsection{Untranscribed speech and unspoken text data}

Figures~\ref{fig:method}(c) and (d) depicts the objectives for the unsupervised data.
For unsupervised data (untranscribed speech and unspoken text), we use the same self-supervised objectives as Maestro \cite{Chen2022MAESTROMS} as
\begin{equation}
    \mathcal{L}_{\text{speech-only}} =
    \lambda_{\text{c}}\mathcal{L}_{\text{c}} + 
    \lambda_{\text{mlm,s}}\mathcal{L}_{\text{mlm,s}}, \qquad
    \mathcal{L}_{\text{text-only}} =
    \lambda_{\text{mlm,t}}\mathcal{L}_{\text{mlm,t}}. \nonumber
\end{equation}
As speaker labels are not available in unspoken text, upsampled text embedding 
is obtained using predicted durations with the speaker embedding for the wildcard identifier \textsf{ANY-SPKR}.

\section{Experiments}\label{sec:eval}

\begin{table*}[tb]
\centering
\caption{TER (\%) and 5-scale SQ in naturalness for different languages. Smaller values are better for TERs whereas larger values are better for SQ.  Values in the bold font indicate the best value.}
\label{tab:objective_all}
\scalebox{0.93}{
\begin{tabular}{l|cccccccc|cccccccc}
\toprule
\multirow{2}{*}{} & \multicolumn{8}{c|}{Seen languages} & \multicolumn{8}{c}{Unseen languages} \\
\toprule
\multirow{2}{*}{}   & \multicolumn{2}{c}{English} & \multicolumn{2}{c}{Spanish} & \multicolumn{2}{c}{Farsi} & \multicolumn{2}{c}{Slovenian} & \multicolumn{2}{|c}{Bulgarian} & \multicolumn{2}{c}{Afrikaans} & \multicolumn{2}{c}{Tamil} & \multicolumn{2}{c}{Turkish} \\
                    & TER        & SQ        & TER        & SQ        & TER        & SQ        & TER        & SQ        & TER        & SQ        & TER        & SQ        & TER        & SQ        & TER        & SQ        \\ \midrule
Tacotron2-G-TTS         & 22.3       & 3.78      & 7.9       & 3.84      & 4.5       & 3.41      & 10.9       & 3.87      & 36.6       & 3.74      & 30.4       & 3.73      & 92.8       & 3.39    & 74.8    & 3.74   \\
MaestroFT-G-TTS & 19.1       & 3.74      & 7.6       & 4.00      & 5.6       & 3.66      & 13.9       & 3.87      & 30.0       & 3.81      & \textbf{24.1}       & 3.68      & 95.2       & 2.62      & 81.9       & 3.99      \\ \midrule
Virtuoso-G-TTS      & 72.7       & 3.48      & 65.0       & 3.67      & 73.0       & 3.40      & 68.6       & 3.60      & 77.0       & 3.54      & 78.2       & 3.39      & 89.9       & 3.46      & 85.7       & 3.59      \\
Virtuoso-G-Pair     & 16.6       & \textbf{4.01}      & 7.1       & \textbf{4.06}      & 4.9       & \textbf{3.85}      & \textbf{6.8}       & \textbf{3.99}      & 23.5       & \textbf{3.83}      & 25.6       & \textbf{4.02}      & 27.4       & \textbf{4.35}      & 38.0       & 4.02      \\
Virtuoso-G-All      & 17.8       & 3.98      & 7.3       & 4.05      & \textbf{4.4}       & 3.77      & 7.3       & 3.93      & 25.6       & \textbf{3.83}      & 28.3       & 3.82      & \textbf{25.0}   & 4.23      & 24.1       & \textbf{4.06}      \\
Virtuoso-G-All-LID  & $\textbf{12.2}$       & 2.93      & 7.0       & 3.18      & 6.4       & 3.20      & 7.0       & 3.32      & 24.0       & 3.46      & 39.1       & 3.02      & 46.5       & 3.26      & \textbf{19.5}       & 3.33      \\
Virtuoso-B-All-LID  & 15.3       & 3.97      & \textbf{6.2}       & 4.05      & 6.9       & 3.82      & 7.0       & 3.92    & \textbf{22.6}   & 3.82  & 28.9     & 3.94     & 29.5       & 4.15      & 20.2       & 4.03      \\ \midrule
Natural             & 8.6       & --      & 5.8       & --      & 3.7       & --      & 17.8       & --      & 5.2       & --  & 12.4       & --      & 16.3       & --      & 5.3       & --      \\
\bottomrule
\end{tabular}
}
\vspace{-2mm}
\end{table*}

\begin{table}[tb]
\centering
\caption{Average TERs and SQ on 10 seen and 4 unseen languages.}
\vspace{0.5mm}
\label{tab:objective_avg}
\begin{tabular}{l|cccc}
\toprule
\multirow{2}{*}{}   & \multicolumn{2}{c}{Seen} & \multicolumn{2}{c}{Unseen} \\
                    & TER          & SQ          & TER             & SQ              \\ \midrule
Tacotron2-G-TTS         & 11.5         & 3.86        & 58.7            & 3.65            \\
MaestroFT-G-TTS & 11.0        & 3.93        & 57.8            & 3.53           \\ \midrule
Virtuoso-G-TTS      & 65.9         & 3.67        & 82.7            & 3.50           \\
Virtuoso-G-Pair     & 9.9         & \textbf{4.05}       & 28.6            & \textbf{4.06}            \\
Virtuoso-G-All      & 10.0        & 4.01        & 25.8            & 3.99            \\
Virtuoso-G-All-LID  & 10.1        & 3.37       & 32.3            & 3.27            \\
Virtuoso-B-All-LID  & \textbf{9.6} & 4.01       & \textbf{25.3}            & 3.99            \\ \midrule
Natural             & 8.6         & --           & 9.8            & --            \\
\bottomrule
\end{tabular}
\vspace{-2mm}
\end{table}

\begin{table}[tb]
\centering
\caption{Subjective 5-scale MOSs in naturalness by human raters. Values in the bold font indicate the best ones.}
\label{tab:subjective}
\vspace{0.5mm}
\begin{tabular}{l|ccc}
\toprule
                    & English               & Spanish               & Tamil               \\ \midrule
Tacotron2-G-TTS         & 3.31$\pm$0.05 & 3.53$\pm$0.09 & 1.59$\pm$0.09 \\
MaestroFT-G-TTS & 3.67$\pm$0.04 & 3.66$\pm$0.07 & 1.24$\pm$0.05 \\ \midrule
Virtuoso-G-TTS      & 1.87$\pm$0.05 & 1.60$\pm$0.10 & 1.28$\pm$0.07 \\
Virtuoso-G-Pair     & \textbf{3.79}$\pm$0.04 & \textbf{3.96}$\pm$0.07 & \textbf{3.39}$\pm$0.08 \\
Virtuoso-G-All      & \textbf{3.81}$\pm$0.04 & \textbf{3.89}$\pm$0.07 & 2.98$\pm$0.08 \\
Virtuoso-G-All-LID  & 1.89$\pm$0.04 & 2.36$\pm$0.08 & 1.89$\pm$0.08 \\
Virtuoso-B-All-LID  & 3.71$\pm$0.04 & \textbf{4.01}$\pm$0.07 & 2.89$\pm$0.08 \\ \bottomrule
\end{tabular}
\vspace{-1mm}
\end{table}

\begin{table}[tb]
\centering
\caption{The TER and SQ of the fine-tuned models using 1 hour (FT-1) and all (FT-All) of the paired TTS data for each language. Virtuoso-G-All was used as a base model for fine-tuning (Pretrain).}
\label{tab:zero}
\scalebox{0.87}{
\begin{tabular}{l|cccccccc}
\toprule
\multirow{2}{*}{} & \multicolumn{2}{c}{Bulgarian} & \multicolumn{2}{c}{Afrikaans} & \multicolumn{2}{c}{Tamil} & \multicolumn{2}{c}{Turkish} \\
                  & TER        & SQ        & TER        & SQ        & TER        & SQ        & TER        & SQ        \\ \midrule
Pretrain    & 25.6       & 3.83      & 28.3       & 3.82      & 25.0   & 4.23      & 24.1       & 4.06      \\
\, + FT-1    & 11.0       & 4.06      & 16.2       & 3.87      & \textbf{18.7}       & \textbf{4.28}  & 8.3       & 3.94      \\
\, + FT-All   & \textbf{7.6}   & \textbf{4.10}      & \textbf{14.8}       & \textbf{3.91}      & 21.1       &4.15      & \textbf{6.4}       & \textbf{3.97}      \\ \midrule
Natural           & 5.2       & -  & 12.4       & --      & 16.3       & --      & 5.3       & --      \\
\bottomrule
\end{tabular}
}
\vspace{-2mm}
\end{table}

\subsection{Experimental conditions}

\subsubsection{Dataset}
\vspace{-1mm}
 A proprietary TTS dataset consisting of 1.5k hours of speech including 40 languages (English, Arabic, Mandarin, Czech, Danish, Spanish, Filipino, Gaelic, Hebrew, Hungarian, Icelandic, Javanese, Latvian, Norwegian, Dutch, Portuguese, Romanian, Russian, Slovakian, Slovenian, Ukrainian, Bengali, Welsh, German, Greek, Estonian, Farsi, Finnish, French, Hindi, Indonesian, Italian, Korean, Lithuanian, Malay, Polish, Serbian, Thai, and Vietnamese) was used as the paired TTS data.
 The total number of speakers in the paired TTS data was 284.
The paired ASR data included VoxPopuli \cite{wang21voxpopuli}, MLS \cite{pratap20mls}, Babel \cite{gales14babel}, and FLEURS \cite{Conneau2022FLEURSFL} following Maestro-U \cite{maestro_u}.
The untranscribed speech data included 429k hours of speech-only data consisting of VoxPopuli, MLS, CommonVoice \cite{ardia19commonvoice}, and Babel as \cite{Chen2022MAESTROMS}.
The unspoken text data contained the VoxPopuli text dataset (3GBytes) and mC4 \cite{xue20mt5} spanning 101 languages (15TBytes) \cite{Chen2022MAESTROMS}.

\vspace{-1mm}
\subsubsection{Model specifications}
\vspace{-1mm}
The specifications of the RNN-T decoder and the speech encoder were the same as those of Maestro \cite{Chen2022MAESTROMS}.
We used 10-layer Conformer blocks for the shared encoder.
We concatenated the text token embedding with a 64-dimensional speaker embedding and then fed it to the text encoder.  The other settings for the text encoder were the same as Maestro.
The speech decoder had 8--headed self-attention blocks with lightweight convolutions, taking a sequence of 1,024-dimensional shared-encoder outputs, the speaker embedding, and the 8-dimensional global reference encoder output.
The target of the speech decoder was a sequence of 80-dimensional mel-spectrogram 
extracted from a speech waveform at 16~kHz sampling (25~ms window length, 10~ms frame shift).
We used UTF-8 bytes~\cite{Li2019BytesAA} for the byte inputs and 6,100 vocabulary size for the grapheme inputs.
We used a WaveGrad neural vocoder with 50 iterations \cite{wavegrad} to reconstruct speech waveform from a predicted mel-spectrogram.

\vspace{-1mm}
\subsubsection{Training specifications}
\vspace{-1mm}
We included the paired ASR, paired TTS, untranscribed speech, and unspoken text data with a fixed effective batch size of (256, 512, 1,024, 2,048) in each batch.
We set the weighting terms $(\lambda_{\text{sd}},\lambda_{\text{rnnt}},\lambda_{\text{c}},\lambda_{\text{mlm},\text{s}},\lambda_{\text{d}},\lambda_{\text{mm}})$ to $(1.0,4.0,1.0,1.0,1.0,0.3)$.
While we set $\lambda_{\text{mlm,t}}$ to $2.0$ without language ids, we upscaled it to $12.0$ when injecting language ids as in Maestro-U~\cite{maestro_u}.
We leveraged curriculum learning as in Maestro, while we started to include paired TTS and ASR data after 300k steps.
We used the same learning rate scheduling and exponential moving average as Maestro.
All the models described in Section~\ref{sec:eval-method} were trained for 500k iterations after starting to include the paired data.
These models were trained for about two weeks using Google Cloud TPUs.

\vspace{-1mm}
\subsubsection{Models}\label{sec:eval-method}
\vspace{-1mm}
We compared baseline models and Virtuoso with different input tokens and training data.
Audio samples are available at \cite{virtuoso_samples}.
In each method name, ``G'' and ``B'' indicate the grapheme and byte input, respectively. 
We trained two baseline models, (1) \textit{\textbf{Tacotron2-G-TTS}}: Tacotron2 \cite{tacotron2} trained on the paired TTS data with a grapheme sequence as input text representation, and
(2) \textit{\textbf{MaestroFT-G-TTS}}: Fine-tuned pretrained Maestro for the TTS task, where alignments between grapheme and speech features were computed using the Maestro's pretrained speech encoder, shared encoder, and RNN-T decoder then the text encoder and speech decoder were fine-tuned with the speech reconstruction loss.

We conducted an ablations study for 
Virtuoso with grapheme-based text representation.
\textit{\textbf{Virtuoso-G-TTS}} only used paired TTS data, \textit{\textbf{Virtuoso-G-Pair}} used paired ASR and TTS data, \textit{\textbf{Virtuoso-G-All}} used all the paired and unpaired data, and \textit{\textbf{Virtuoso-G-All-LID}} introduced language IDs and a language adapter as in Maestro-U \cite{maestro_u}.
Finally, \textit{\textbf{Virtuoso-B-All-LID}} used the UTF-8 bytes rather than graphemes as its text representation (like Byte2Speech \cite{He2021MultilingualBM}) with all the paired and unpaired data plus language IDs and a language adapter like \textit{\textbf{Virtuoso-G-All-LID}}. 

\vspace{-1mm}
\subsubsection{Evaluation metrics}
\vspace{-1mm}
We evaluated the models with three metrics.
Since the models output speech features directly from graphemes or byte sequences, we used token error rates (TER) to evaluate the accuracy of linguistic content in synthetic speech.
We used a multilingual ASR model trained on data that did not contain the paired TTS data but included all the languages in the evaluation.
Subjective listening evaluations in naturalness using 5-scale mean opinion score (MOS) tests for three languages (English, Spanish, Tamil) were conducted to evaluate the synthetic speech.
As it is difficult to have enough raters for some low-resource languages, we also used an automatically computed 5-scale MOS in naturalness by SQuId \cite{SQuId} (SQ). 
SQ doesn't map perfectly to subjective MOS and is less sensitive to linguistic correctness since the model has largely seen ratings for high-quality TTS samples (ranging between 3.0 and 5.0).  However, it is still useful for relative comparisons between models within the same language.

\subsection{Experimental results}

\subsubsection{Seen languages}
\vspace{-1mm}
Among languages in the paired TTS data, we selected English, Spanish, Farsi, and Slovenian as ``seen'' languages for the evaluation.
We selected one speaker for each language and used them in the evaluation.
The left part of Table~\ref{tab:objective_all} shows the experimental results for seen languages\footnote{TER for natural speech was worse than that of synthetic speech in Slovenian.  This can be due to large variations in the natural speech. }.
It can be seen from the TERs in the table that Virtuoso-G-TTS was significantly less intelligible than other models. This suggests that the paired TTS data was not enough to train a large-scale Virtuoso model.
On the other hand, Virtuoso-G-Pair, which uses both TTS and ASR paired data, achieved better SQ and TER than the baseline models for all the seen languages.
Introducing unsupervised data in addition to the supervised one had small or no impact both in SQ and TER.
Among all models, Virtuoso-G-Pair consistently achieved the best SQ.

Table~\ref{tab:objective_avg} also gives average TER and SQ over ten seen languages.
Like the four seen languages in Table~\ref{tab:objective_all}, Virtuoso-G-Pair achieved the highest SQ.
As the Virtuoso-G-Pair models had more paired TTS data in a mini-batch than Virtuoso-G-All, the reconstruction loss could be smaller at the same number of training steps.
The low SQ for Virtuoso-G-All-LID can be due to the larger loss weight for text-only data as in Maestro-U \cite{maestro_u}.
This could lead to the higher speech reconstruction loss values in the TTS learning. 
Further investigation and tuning of the training configurations is a future work.

\vspace{-1mm}
\subsubsection{Unseen languages}
\vspace{-1mm}
We selected Bulgarian, Afrikaans, Tamil, and Turkish as ``unseen'' languages.
Note that the paired TTS data did not include these ``unseen'' languages whereas the paired ASR data and unsupervised data included them (e.g., FLEURS \cite{Conneau2022FLEURSFL}, mC4 \cite{xue20mt5}).
As there was no speaker for these languages in the paired TTS data, we selected one speaker from a similar seen language for each unseen language (Russian for Bulgarian, Dutch for Afrikaans, Hindi for Tamil, and French for Turkish) for evaluation.
We found that using speaker embedding from a similar language gave better performance.

The right part of Table~\ref{tab:objective_all} shows the results for unseen languages.
While the baseline models performed relatively well for Bulgarian and Afrikaans, they completely failed to synthesize intelligible speech for Tamil and Turkish.
This can be because the input tokens for these two languages were significantly different from the seen languages in the paired TTS data.
On the other hand, Virtuoso models with the paired ASR data achieved decent performance even for these unseen languages, as the additional paired ASR and unpaired data can provide some signals about these input tokens.

Table~\ref{tab:objective_avg} shows that Virtuoso models with the unpaired data significantly improved TER over those without it, demonstrating that the introduction of the unpaired data can improve the linguistic accuracy in unseen languages.  Like the seen languages, Virtuoso-B-All-LID achieved the lowest TER, while Virtuoso-G-Pair got the highest SQ.

As SQ is less sensitive to linguistic correctness \cite{SQuId}, we also conducted subjective 5-scale MOS by human raters.
Table~\ref{tab:subjective} lists the MOS test results in English (seen), Spanish (seen), and Tamil (unseen).
Although there are inconsistency in absolute scores between Tables~\ref{tab:objective_all} and \ref{tab:subjective}, their rankings are somewhat consistent between them. 
We can see that the best Virtuoso model showed encouraging MOS of 3.39 for the unseen language.


Finally, we conducted an experiment to fine-tune a pretrained Virtuoso-G-All model on unseen languages.
During fine-tuning, all self-supervised losses were disabled; only ASR and reconstruction losses were used.
Table~\ref{tab:zero} gives the experimental results.
We can see that fine-tuning significantly improved TER even with 1~hour paired TTS data whereas SQ was less affected.

\vspace{-1mm}
\section{conclusions}
\vspace{-1mm}
This paper presented \emph{Virtuoso}, a massively multilingual joint speech--text semi-supervised learning framework for TTS.
Multilingual TTS models can be trained using both supervised (paired ASR and TTS data) and unsupervised (untranscribed speech and unspoken text) data including hundreds of languages, by extending \emph{Maestro} to synthetic tasks.
Experimental results demonstrated that the multilingual TTS models achieved significantly more intelligible and natural synthetic speech than baseline grapheme-based Tacotron2 and fine-tuned Maestro models.
We also demonstrated its capability to synthesize speech in languages without paired TTS data.
It has a potential to greatly increase the language coverage in multilingual TTS using unpaired speech and text data.

Future work includes exploring more efficient ways to inject unpaired data and improving the quality. 
Training a multilingual model from the data in hundreds of languages is also future work.

\newpage
\bibliographystyle{IEEEbib}
\footnotesize 
\bibliography{refs}

\begin{thebibliography}{10}

\bibitem{Zen_SLF_TASLP}
H.~Zen, N.~Braunschweiler, S.~Buchholz, et~al.,
\newblock ``Statistical parametric speech synthesis based on speaker and
  language factorization,''
\newblock {\em IEEE Trans. Audio Speech Lang. Process.}, vol. 20, no. 6, pp.
  1713--1724, 2012.

\bibitem{Li2016MultiLanguageMA}
B.~Li and H.~Zen,
\newblock ``Multi-language multi-speaker acoustic modeling for {LSTM-RNN} based
  statistical parametric speech synthesis,''
\newblock in {\em Proc. Interspeech}, 2016.

\bibitem{He2021MultilingualBM}
M.~He, J.~Yang, L.~He, et~al.,
\newblock ``Multilingual {Byte2Speech} models for scalable low-resource speech
  synthesis,''
\newblock {\em arXiv:2103.03541}, 2021.

\bibitem{Conneau22xlsr}
A.~Conneau, A.~Baevski, R.~Collobert, et~al.,
\newblock ``Unsupervised cross-lingual representation learning for speech
  recognition,''
\newblock in {\em Proc. Interspeech}, 2021, pp. 2426–--2430.

\bibitem{Bapna22mslam}
A.~Bapna, C.~Cherry, Y.~Zhang, et~al.,
\newblock ``{mSLAM}: {M}assively multilingual joint pre-training for speech and
  text,''
\newblock {\em arXiv:2202.01374}, 2022.

\bibitem{Chen2022MAESTROMS}
Z.~Chen, Y.~Zhang, A.~Rosenberg, et~al.,
\newblock ``{MAESTRO}: {M}atched speech text representations through modality
  matching,''
\newblock in {\em Proc. Interspeech}, 2022, pp. 4093--4097.

\bibitem{Baevski2020wav2vec2A}
A.~Baevski, H.~Zhou, A.-R. Mohamed, et~al.,
\newblock ``{wav2vec 2.0}: {A} framework for self-supervised learning of speech
  representations,''
\newblock {\em arXiv:2006.11477}, 2020.

\bibitem{Chung2021w2vBERTCC}
Y.-A. Chung, Y.~Zhang, W.~Han, et~al.,
\newblock ``{w2v-BERT}: {C}ombining contrastive learning and masked language
  modeling for self-supervised speech pre-training,''
\newblock {\em Proc. ASRU}, pp. 244--250, 2021.

\bibitem{Chen2022WavLMLS}
S.~Chen, C.~Wang, Z.~Chen, et~al.,
\newblock ``{WavLM}: {L}arge-scale self-supervised pre-training for full stack
  speech processing,''
\newblock {\em arXiv:2110.13900}, 2022.

\bibitem{chen21injecting}
Z.~Chen, Y.~Zhang, A.~Rosenberg, et~al.,
\newblock ``Injecting text in self-supervised speech pretraining,''
\newblock in {\em Proc. ASRU}, 2021, pp. 251--258.

\bibitem{tang-etal-2022-unified}
Y.~Tang, H.~Gong, N.~Dong, et~al.,
\newblock ``Unified speech-text pre-training for speech translation and
  recognition,''
\newblock in {\em Proc. ACL}, 2022, pp. 1488--1499.

\bibitem{pmlr-v162-bai22d}
H.~Bai, R.~Zheng, J.~Chen, et~al.,
\newblock ``{A}$^3${T}: Alignment-aware acoustic and text pretraining for
  speech synthesis and editing,''
\newblock in {\em Proc. ICML}, 2022, pp. 1399--1411.

\bibitem{ao-etal-2022-speecht5}
J.~Ao, R.~Wang, L.~Zhou, et~al.,
\newblock ``{S}peech{T}5: Unified-modal encoder-decoder pre-training for spoken
  language processing,''
\newblock in {\em Proc. ACL}, 2022, pp. 5723--5738.

\bibitem{Karita_joint_ASR_TTS}
S.~Karita, S.~Watanabe, T.~Iwata, et~al.,
\newblock ``Semi-supervised end-to-end speech recognition using text-to-speech
  and autoencoders,''
\newblock in {\em Proc. ICASSP}, 2019, pp. 6166--6170.

\bibitem{Wang2020ImprovingSR}
G.~Wang, A.~Rosenberg, Z.~Chen, et~al.,
\newblock ``Improving speech recognition using consistent predictions on
  synthesized speech,''
\newblock {\em Proc. ICASSP}, pp. 7029--7033, 2020.

\bibitem{Lim2020JDITJT}
D.~Lim, W.~Jang, G.~O, et~al.,
\newblock ``{JDI-T}: {J}ointly trained duration informed {Transformer} for
  text-to-speech without explicit alignment,''
\newblock in {\em Proc. Interspeech}, 2020, pp. 4004--4008.

\bibitem{Ren2019AlmostUT}
Y.~Ren, X.~Tan, T.~Qin, et~al.,
\newblock ``Almost unsupervised text to speech and automatic speech
  recognition,''
\newblock {\em arXiv:1905.06791}, 2019.

\bibitem{makishima22_interspeech}
N.~Makishima, S.~Suzuki, A.~Ando, et~al.,
\newblock ``Speaker consistency loss and step-wise optimization for
  semi-supervised joint training of {TTS} and {ASR} using unpaired text data,''
\newblock in {\em Proc. Interspeech}, 2022, pp. 526--530.

\bibitem{oshaughnessy-1998-multilingual}
D.~O{'}Shaughnessy,
\newblock ``Multilingual text-to-speech synthesis: {T}he {Bell} labs
  approach,''
\newblock {\em Computational Linguistics}, vol. 24, no. 4, 1998.

\bibitem{Zhang2019LearningTS}
Y.~Zhang, R.~J. Weiss, H.~Zen, et~al.,
\newblock ``Learning to speak fluently in a foreign language: {M}ultilingual
  speech synthesis and cross-language voice cloning,''
\newblock in {\em Proc. Interspeech}, 2019, pp. 2080--2084.

\bibitem{Zhao2020bilingual}
S.~Zhao, T.~H. Nguyen, H.~Wang, et~al.,
\newblock ``Towards natural bilingual and code-switched speech synthesis based
  on mix of monolingual recordings and cross-lingual voice conversion,''
\newblock {\em Proc. Interspeech}, pp. 2927--2931, 2020.

\bibitem{Yang2022Lifelong}
M.~Yang, S.~Ding, T.~Chen, et~al.,
\newblock ``Towards lifelong learning of multilingual text-to-speech
  synthesis,''
\newblock in {\em Proc. ICASSP}, 2022, pp. 8022--8026.

\bibitem{prakash19_ssw}
A.~Prakash, A.~L. {Thomas}, S.~Umesh, et~al.,
\newblock ``{Building Multilingual End-to-End Speech Synthesisers for Indian
  Languages},''
\newblock in {\em Proc. SSW}, 2019, pp. 194--199.

\bibitem{ogayo22_interspeech}
P.~Ogayo, G.~Neubig, and A.~{W Black},
\newblock ``Building {African} voices,''
\newblock in {\em Proc. Interspeech}, 2022, pp. 1263--1267.

\bibitem{Li2019BytesAA}
B.~Li, Y.~Zhang, T.~N. Sainath, et~al.,
\newblock ``Bytes are all you need: {E}nd-to-end multilingual speech
  recognition and synthesis with bytes,''
\newblock {\em Proc. ICASSP}, pp. 5621--5625, 2019.

\bibitem{elias2021parallel}
I.~Elias, H.~Zen, J.~Shen, et~al.,
\newblock ``Parallel tacotron: Non-autoregressive and controllable tts,''
\newblock in {\em Proc. ICASSP}, 2021, pp. 5709--5713.

\bibitem{hsu2018hierarchical}
W.-N. Hsu, Y.~Zhang, R.~Weiss, et~al.,
\newblock ``Hierarchical generative modeling for controllable speech
  synthesis,''
\newblock in {\em Proc. ICLR}, 2019.

\bibitem{Yamagishi2019CSTRVC}
J.~Yamagishi, C.~Veaux, and K.~MacDonald,
\newblock ``{CSTR VCTK} corpus: {English} multi-speaker corpus for {CSTR} voice
  cloning toolkit (version 0.92),'' 2019.

\bibitem{He2022MaskedAA}
K.~He, X.~Chen, S.~Xie, et~al.,
\newblock ``Masked autoencoders are scalable vision learners,''
\newblock in {\em Proc. CVPR}, 2022, pp. 15979--15988.

\bibitem{wang21voxpopuli}
C.~Wang, M.~Riviere, A.~Lee, et~al.,
\newblock ``{VoxPopuli}: {A} large-scale multilingual speech corpus for
  representation learning, semi-supervised learning and interpretation,''
\newblock {\em arXiv:2101.00390}, 2021.

\bibitem{pratap20mls}
V.~Pratap, Q.~Xu, A.~Sriram, et~al.,
\newblock ``{MLS}: {A} large-scale multilingual dataset for speech research,''
\newblock {\em arXiv:2012.03411}, 2019.

\bibitem{gales14babel}
M.~J. Gales, K.~M. Knill, A.~Ragni, et~al.,
\newblock ``Speech recognition and keyword spotting for low-resource languages:
  Babel project research at {CUED},''
\newblock in {\em Proc. SLTU}, 2014, pp. 16–--23.

\bibitem{Conneau2022FLEURSFL}
A.~Conneau, M.~Ma, S.~Khanuja, et~al.,
\newblock ``{FLEURS}: {F}ew-shot learning evaluation of universal
  representations of speech,''
\newblock {\em arXiv:2205.12446}, 2022.

\bibitem{maestro_u}
Z.~Chen, A.~Bapna, A.~Rosenberg, et~al.,
\newblock ``{Maestro-U:} leveraging joint speech--text representation learning
  for zero supervised speech {ASR},''
\newblock {\em arXiv:2210.10027}, 2022.

\bibitem{ardia19commonvoice}
R.~Ardila, M.~Branson, K.~Davis, et~al.,
\newblock ``{Common Voice}: {A} massively-multilingual speech corpus,''
\newblock {\em arXiv:1912.06670}, 2019.

\bibitem{xue20mt5}
L.~Xue, N.~Constant, A.~Roberts, et~al.,
\newblock ``{mT5}: {A} massively multilingual pre-trained text-to-text
  transformer,''
\newblock {\em arXiv:2010.11934}, 2020.

\bibitem{wavegrad}
N.~Chen, Y.~Zhang, H.~Zen, et~al.,
\newblock ``{WaveGrad}: {Estimating} gradients for waveform generation,''
\newblock in {\em Proc. ICLR}, 2021.

\bibitem{virtuoso_samples}
``Audio samples for {Virtuoso},''
  \url{https://google.github.io/tacotron/publications/virtuoso/}.

\bibitem{tacotron2}
J.~Shen, R.~Pang, R.~J. Weiss, et~al.,
\newblock ``Natural {TTS} synthesis by conditioning {WaveNet} on mel
  spectrogram predictions,''
\newblock in {\em Proc. ICASSP}, 2018, pp. 4779--4783.

\bibitem{SQuId}
T.~Sellam, A.~Bapna, J.~Camp, et~al.,
\newblock ``{SQuId}: Measuring speech naturalness in many languages,''
\newblock {\em arXiv:2210.06324}, 2022.

\end{thebibliography}

\end{document}